\documentclass[twocolumn,prl,floatfix,superscriptaddress,showpacs]{revtex4}
\usepackage{graphicx}
\usepackage{latexsym}
\usepackage{amssymb}

\begin{document}

\title{Novel metallic and insulating states at a bent quantum Hall junction}

\author{M. Grayson}
\affiliation{Walter Schottky Institut, Technische Universit\"at
M\"unchen, 85748 Garching, Germany}
\affiliation{Max-Planck-Institut f\"ur Festk\"orperforschung, 70569 Stuttgart, Germany}
\author{L. Steinke}
\affiliation{Walter Schottky Institut, Technische Universit\"at
M\"unchen, 85748 Garching, Germany}
\author{D. Schuh}
\altaffiliation[Present address: ]{Universit\"at Regensburg, Institut f\"ur Experimentelle und Angewandte Physik, 93040 Regensburg, Germany.}
\affiliation{Walter Schottky Institut, Technische Universit\"at
M\"unchen, 85748 Garching, Germany}
\author{M. Bichler}
\affiliation{Walter Schottky Institut, Technische Universit\"at
M\"unchen, 85748 Garching, Germany}
\author{L. Hoeppel}
\affiliation{Max-Planck-Institut f\"ur Festk\"orperforschung, 70569 Stuttgart, Germany}
\author{J. Smet}
\affiliation{Max-Planck-Institut f\"ur Festk\"orperforschung, 70569 Stuttgart, Germany}
\author{K. v. Klitzing}
\affiliation{Max-Planck-Institut f\"ur Festk\"orperforschung, 70569 Stuttgart, Germany}
\author{D. K. Maude}
\affiliation{Grenoble High Magnetic Field Laboratories CNRS, 38042 Grenoble, France}
\author{G. Abstreiter}
\affiliation{Walter Schottky Institut, Technische Universit\"at
M\"unchen, 85748 Garching, Germany}

\date{\today}

\begin{abstract}
A non-planar geometry for the quantum Hall (QH) effect is studied, whereby two quantum Hall (QH) systems are joined at a sharp right angle.  When both facets are at equal filling factor $\nu$ the junction hosts a channel with non-quantized conductance, dependent on $\nu$.  The state is metallic at $\nu = 1/3$, with conductance along the junction increasing as the temperature $T$ drops. At $\nu = 1, 2$ it is strongly insulating, and at $\nu = 3, 4$ shows only weak $T$ dependence.  Upon applying a dc voltage bias along the junction, the differential conductance again shows three different behaviors.  Hartree calculations of the dispersion at the junction illustrate possible explanations, and differences from planar QH structures are highlighted.
\end{abstract}

\maketitle

Experimental studies of reduced dimensional conductors are relevant for both nanoelectronics and basic physics.  Conductance in semiconductor nanowires \cite{yac97,kau99,fie90,tar95} is limited by disorder and interactions which can backscatter propagating charge.  In chiral one-dimensional (1D) systems like quantum Hall (QH) edges \cite{hal82,wen90} charge propagates in only one direction but can be tailored to backscatter and interact at one \cite{rod03} or several \cite{ji03} point-like constrictions.  More about 1D systems could be learned if two QH edges could be coupled to an extended disorder potential to reconstitute a wire.  In planar geometries, however, lithographically defined edges \cite{hau88} have soft confinement potentials and suffer from edge reconstruction \cite{chk92,cha94}, and sharp confinement structures which include a tunnel barrier \cite{kan00} spatially separate forward and reverse movers, reducing both interactions and backscattering.  

Here we study a conducting state at the corner of two QH systems joined at a 90$^{\mathrm o}$ angle, where the corner geometry itself serves as a sharp QH boundary for both facets.  The corner junction hosts a conducting channel which carries current across the full macroscopic length of the sample, and whose conductance is not quantized even though the facets are in QH states.  At different filling factor $\nu$, both metallic and insulating  conductance along the junction are observed as a function of temperature and voltage. The length-dependent conductance for some $\nu$ suggest disorder mediated scattering.  Most striking is the metallic behavior at $\nu = 1/3$ whereby conductance along the junction increases with decreasing temperature.  Such non-planar confinement structures are unconventional for the QH effect, and Hartree calculations of a sharp corner illustrate the expected dispersions, clarifying possible origins of the observed phases. 
\begin{figure}[!ht]
\center \includegraphics[width=8.5cm]{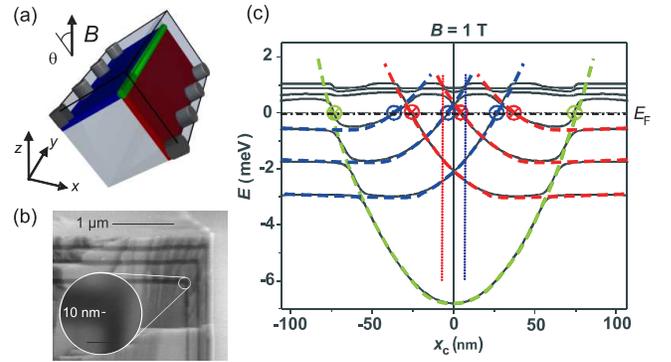}
\caption{(a) Schematic of the bent quantum well subjected to a quantizing $B$ field. The electrons in the facets (blue and red) and corner accumulation wire (green) are colored according to their dispersions in panel (c). (b) Scanning electron micrograph of a diagnostic structure with AlAs (dark) and GaAs (light) bands.  The corner curvature $R$ is sharper than the 10 nm resolution limit. (c) Hartree calculations of the dispersion at a sharp corner (black) from Eq. \ref{eq:Hartree}, overlaid with left- (blue) and right-facing (red) sharp QH edge dispersions and an accumultion wire dispersion (green).  Vertical dotted lines represent the effective hard walls seen by the edge states.}
\label{fig:sampledispers}
\end{figure}

The bent quantum well is fabricated by epitaxially overgrowing a GaAs/AlGaAs heterostructure on a cleaved corner \cite{gra05} [Fig.~\ref{fig:sampledispers}(a,b)]. A GaAs well layer is topped with an Al$_{0.7}$Ga$_{0.3}$As barrier with modulation doping at a distance $d$ = 120 nm, forming an L-shaped heterointerface where electrons are confined.  The facets have near-equal densities $n_1 = 1.10   \times 10^{11}~{\rm cm}^{-2}$ and $n_2 = 1.28 \times 10^{11}~{\rm cm}^{-2}$, and a junction length $L$ = 2 mm for sample A ($n_1 = 1.11 \times 10^{11}~{\rm cm}^{-2}, n_2 = 1.45 \times 10^{11}~{\rm cm}^{-2}$, and $L$ = 4.5 mm for sample B) \cite{sample}  with a mobility estimated at around $\mu \sim 5 \times 10^5~{\rm cm}^2$/Vs.  Additional samples showed the same behavior.

\begin{figure}[t]
\center \includegraphics[width=8.5cm]{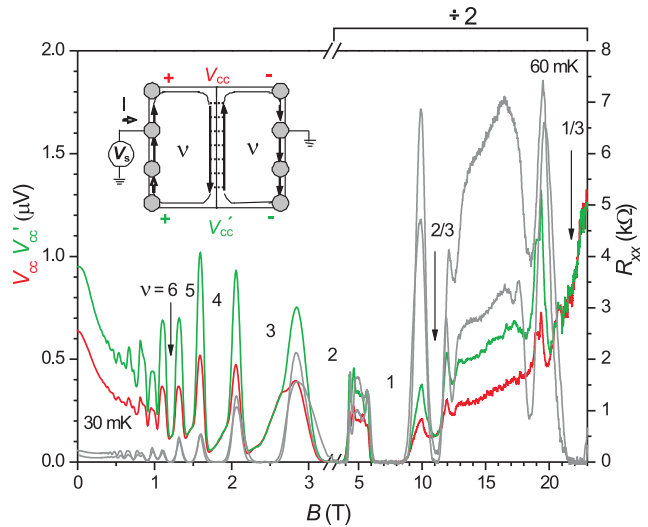}
\caption{(Sample A) Longitudinal resistance $R_{xx}$ within each facet (grey), and cross-corner voltages $V_{\rm cc}$ (red) and $V'_{\rm cc}$ (green).  Non-zero $V_{\rm cc}$ minima indicate finite conduction along the junction. Inset: schematic of edge states and backscattering of current along the junction.}
\label{fig:rxxrcc}
\end{figure}

A tilted magnetic field $B$ at angle $\theta$ can induce the QH effect in both facets.  At a conventional QH edge, the perpendicular field component $B_\perp$ induces a mobility gap within each facet, leaving chiral 1D edge channels to carry current at the periphery \cite{hal82,wen90} (Fig.~\ref{fig:rxxrcc}, inset).  The most prominent gapped states occur when the filling factor $\nu = hn/eB_\perp$  is an integer or odd-denominator fraction, and for the integer QH effect, $\nu$ also counts the number of 1D edge channels. This Letter considers only equal $\nu$ on both facets (for other $B$ orientations and $\nu$ ratios, see Ref. \cite{gra05}.)

To understand what sort of edge states may exist at the bent QH junction, we first calculate the dispersion at finite $B$ using the Hartree potential $V_{\rm H} (x,z)$ solved at $B = 0$ for a sharp corner :
\begin{equation}
\left\{ \frac{({\bf p} + e{\bf A})^2}{2m^*} + V_{\rm H}(x,z)\right\} \psi = E \psi 
\label{eq:Hartree}
\end{equation}
\noindent Fig.~\ref{fig:sampledispers}(a) shows the Cartesian coordinates, and the calculations assume equal charge density on both facets, $n_1 = n_2 = 1 \times 10^{11}~{\rm cm}^{-2}$, neglecting spin for simplicity.  By choosing the Landau gauge ${\bf A} = (0,Bx,0)$, momentum $k_y$ is a good quantum number, and the dispersion $E_m(k_y)$ results from the eigenvalue problem of Eq.~\ref{eq:Hartree} for $\psi_{m,k_y} = \phi_m(x,z)e^{ik_yy}$, where $m$ is the Landau index.

Fig.~\ref{fig:sampledispers}(c) shows the dispersion (black) versus projected orbit centre $x_c = k_yl_B^2$ for the lowest energy levels ($l_B = \sqrt{\hbar/eB}$ is the magnetic length.)  For comparison, the dispersion of a right-facing sharp QH edge \cite{hub05} (blue) is shown for a hard wall positioned at the vertical dotted-blue line, mirrored by the red dispersion of a left-facing edge.  These two hard-wall-like dispersions arise because the sudden 90$^{\mathrm o}$ bend in the heterojunction serves as a hard wall for the incident skipping orbits within the opposing facet.  Unlike the planar antiwire of Ref. \cite{kan00}, there is no tunnel barrier separating the two systems, and the edge states from the two orthogonal facets interpenetrate at the corner.  The third subsystem is a deeply bound wire seen previously in Hartree calculations as a 1D accumulation of charge at the corner \cite{gra05} and indicated here by the green parabolic dispersion. This accumulation wire adds two spin-degenerate 1D modes to the edge in each direction and serves as an additional 1D channel for scattering. Together with the 1D edge modes from the QH systems, the model thus predicts as many as $N = \nu + 2$ 1D modes in each direction, whose dispersions anticross in the Hartree solution.

The Hartree calculations apply to the experiment as long as the corner curvature $R$ at the junction is sharper than two quantum lengths: the Fermi wavelength $\lambda_F = 70$ nm and $l_B$.  In Fig.~\ref{fig:sampledispers}(b), $R$ is determined within the resolution of the electron microscope to be $R < 10$ nm. Because $2R < \lambda_F/2$, a 1D accumulation wire should exist at the junction with only a single-subband occupied.  The condition $R = l_B$ determines the field $B_c = \hbar/e R^2 = 6.6$ T below which the $B = 0$ Hartree potential safely approximates the finite $B$ solution.  The condition $B \lesssim B_c$ corresponds to $\nu \gtrsim 1$ in this sample, so the calculations should assist in understanding the integer QH regime.

We now experimentally investigate the bent QH junction.  Zero-resistance minima in the longitudinal resistance $R_{xx}$ for both facets in Fig.~\ref{fig:rxxrcc} (grey) identify the well-formed QH states:  $\nu$ = 1/3, 2/3, 1, 2, 3, 4, 5, 6.  The conductance along the junction is measured using the following 4-point geometry \cite{ren95,kan97}:  a current is driven across the corner to ground with an applied bias $V_{\rm s}$, and the resultant voltage $V_{\rm cc}$ or $V'_{\rm cc}$ is measured between two contacts, one on each facet (Fig.~\ref{fig:rxxrcc}, inset).  

The cross-corner voltage $V_{\rm cc}$ is plotted in red and $V'_{\rm cc}$ in green in Fig.~\ref{fig:rxxrcc}.  With the facets in a QH state, the current along the junction can be calculated \cite{but88} $I_{\rm jct} =  (\nu e^2/h) V_{\rm cc}  =(\nu e^2/h) V'_{\rm cc} $.  Whenever $R_{xx} = 0$, $V_{\rm cc} = V'_{\rm cc}$ represents current conservation entering and exiting the junction. The conductance $G$ along the junction is
\begin{equation}
G = \frac{I_{\rm jct}}{V_{\rm s}} = \nu \frac{e^2}{h}\frac{V_{\rm cc}}{V_{\rm s}} = \nu \frac{e^2}{h}\frac{V'_{\rm cc}}{V_{\rm s}} 
\label{eq:G}
\end{equation}

\begin{figure}[t]
\center \includegraphics[width=8.5cm]{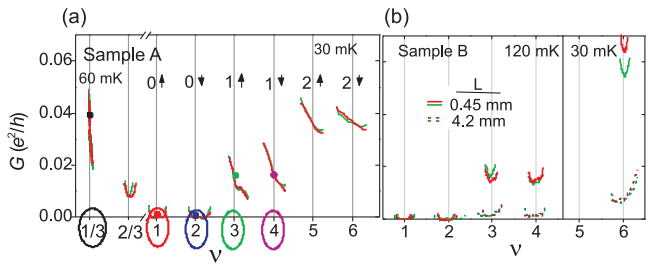}
\caption{Bent QH junction conductance $G$ vs.~$\nu$ (a) (Sample A) Landau index $m$ and spin quantum number $\sigma$ are indicated above each integer $\nu$ revealing a pairwise similarity. Circled $\nu$ are measured as a function of temperature and voltage in Fig. ~\ref{fig:Pdep}(a) and (b). (b) (Sample B) Length dependence of the bent QH junction conductance $G$.  The $L_1$ = 4.2 mm corner-junction was first characterized (red and green, dotted lines), and then scribed to a length $L_2$ = 0.45 mm and remeasured (red and green, solid lines).}
\label{fig:lengthdep}
\end{figure}

The junction conductance for Sample A is plotted versus $\nu$ in Fig.~\ref{fig:lengthdep}(a) for filling factors where $R_{xx} = 0$.  At 30~mK, the junction does not conduct for either $\nu$ = 1 or 2.  At $\nu$ = 3, 4, 5, and 6 the junction conductance falls within the range 0.01- 0.04 of $e^2/h$.  The fractional QH effect $\nu$ = 2/3 shows a similar conductance, while the conductance at $\nu$ = 1/3 at 60 mK is slightly higher.  In all cases, the small and non-quantized junction conductance $G \ll e^2/h$ indicates that charge is strongly backscattered within the junction region.

In Fig. \ref{fig:lengthdep}(a), integer $\nu$ are labelled with both the Landau index $m$ as well as spin $\sigma$.  In diffusive 1D systems, one expects a stepwise increase in conductance with each additional mode, and in Fig.~\ref{fig:lengthdep} such steps are observed to occur pairwise in $\nu$: $\nu$ = 1, 2 ($m$ = 0); 3, 4 ($m$ = 1); and 5, 6 ($m$ = 2).  The conductance thus behaves as though the Landau index $m$, not $\nu$, counts the modes at the junction.  A suppression of spin splitting at this sharp QH junction could explain this result.  A similar lack of spin-splitting at a sharp edge was already experimentally observed in tunneling experiments of sharp-edge systems \cite{hub05} and deserves further scrutiny.

The length dependence $L$ of the conductance for Sample B is shown in Fig.~\ref{fig:lengthdep}(b).  At $\nu$ = 3, 4, and 6, where the dependence could be measured, the short junction ($L_2$ = 0.45 mm, solid lines) conducts better than the long junction ($L_1$ = 4.2 mm, dotted lines), with conductance scaling approximately as $G \sim 1/L$.  If backscattering is distributed uniformly along the length of the junction, 1D conductance can be written $G = (e^2/h) l_0N/L$ where $l_0N$ is the mean free path times the number of modes.  These results suggests mean-free path $l_0N= 7 \mu$m ($\nu$ = 3, 4) and $l_0N = 27 \mu$m ($\nu$ = 6) at the temperatures shown, and provide evidence that the charge backscattering is distributed along the junction.

\begin{figure}[t]
\center \includegraphics[width=8.5cm]{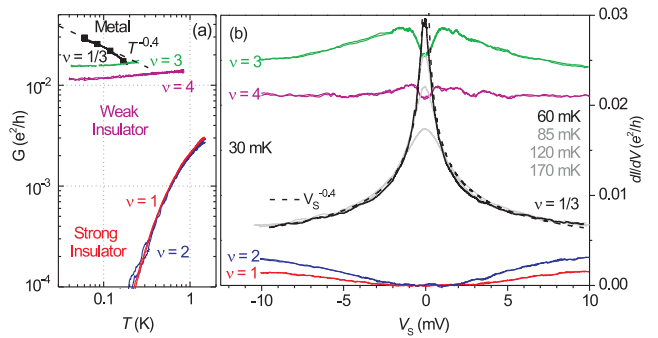}
\caption{Sample A: (a) The temperature dependence of the bent QH junction conductance $G$ vs.~$T$ at the $\nu$ circled in Fig.~\ref{fig:lengthdep}. $\nu$ = 1, 2 are strongly insulating; $\nu$ = 3, 4 weakly insulating; and the fractional $\nu$ = 1/3 metallic. (b) The dc voltage dependence $V_{\rm s}$ of the differential conductance $dI/dV$ for the same $\nu$.  For $\nu$ = 1/3 four temperatures are shown, whose peak values at $V_{\mathrm s} = 0$ correspond to the black squares in panel (a).  Power laws $T^{-0.4}$ and $V^{-0.4}$ plotted for comparison (dotted line).}
\label{fig:Pdep}
\end{figure}

The junction conductance was also measured as a function of temperature $T$ and dc voltage bias $V_{\rm s}$.  Fig.~\ref{fig:Pdep}(a) shows the $T$ dependence of the conductance for $\nu$ = 1/3, 1, 2, 3, and 4.  For each $\nu$, the same behaviour occurs across the entire minimum.  With decreasing $T$, the conductivity along the corner junction either decreases ($\nu$ = 1, 2), stays roughly constant ($\nu$ = 3, 4), or increases ($\nu$ = 1/3), illustrating what we label as strongly insulating, weakly insulating or metallic behaviour, respectively.  In Fig.~\ref{fig:Pdep}(b), the differential conductance dI/dV of the corner QH junction is plotted for the same $\nu$ as a function of $V_{\rm s}$. The insulator $dI/dV$ drops drastically with reduced bias ($\nu$ = 1, 2), whereas the metallic state $dI/dV$ increases, forming a cusp at zero bias ($\nu$ = 1/3).  Varying the temperature up to 170~mK while measuring this cusp shows that most of the temperature dependence occurs at the small biases.  The weakly insulating phase ($\nu$ = 3, 4) shows the weakest bias dependence, with a mild dip at low bias indicating an insulator.  

Possible explanations of these phases must be consistent with the experimentally measured sharp junction curvature of Fig.~\ref{fig:sampledispers}(b).  We therefore base our discussion on the dispersions of Fig.~\ref{fig:sampledispers}(c).  The explanation for the insulator at $\nu$ = 1, 2 is twofold.  It can arise either from an anticrossing band insulator at the corner or from localization of 1D states.  Gaps in the dispersion arise whenever the bands of Fig.~\ref{fig:sampledispers}(c) anticross, and if the Fermi level sits within such a gap, the junction would host a band insulator.  At higher $B$ relevant for $\nu$= 1, 2, the anticrossing gaps from Eq.~\ref{eq:Hartree} increase (results not shown), and are more likely to result in such a band insulator.  Alternately, the $\nu = 1, 2$ insulator could arise according to the scaling theory of localization for 1D systems, since all 1D systems are expected to become insulators in the presence of disorder \cite{abr79} and repulsive interactions \cite{gia88,gor05}.  The limited temperature range of the data in Fig.~\ref{fig:Pdep}(a,b) is insufficient to identify which of these two mechanisms may be responsible, though we note that the conductance does drop faster than a power-law, consistent with both explanations.  We also note that the $\nu$ = 1, 2 temperature dependences perfectly overlap, suggesting a common mechanism.

The weakly insulating behaviour at $\nu$ = 3, 4 may be related to weak localization.  Examining the voltage dependence in Fig.~\ref{fig:Pdep}(b), the zero-bias dip in dI/dV suggests a crossover from a metal to a weakly insulating state below $V_s \sim 1$ mV, which would represent an energy scale for the weak localization.  A careful modelling of the multimode 1D conductance of Fig.~\ref{fig:sampledispers}(c) will be the first step towards identifying these energy scales in the model, and promises to be an interesting subject of future work.

Perhaps most intriguing is the metallic behavior at $\nu$ = 1/3, with a junction conductance that increases as temperature is lowered.  At such high fields $B > B_c$, the Hartree dispersions from Eq. 1 would have to be calculated self-consistently at finite $B$, and must include interactions to correctly account for the Laughlin ground state in the facets.  Though such calculations are beyond the scope of this paper, qualitatively one expects a mixing of the accumulation wire magnetosubband dispersion for electrons \cite{wee88} with the fractional QH edge dispersions for quasiparticles \cite{wen90}.  

Looking at the $\nu = 1/3$ voltage bias curves in Fig.~\ref{fig:Pdep}, it is clear that the conductance is strongly temperature dependent at extremely low temperatures.  The likeliest candidate for such low-energy scattering is electron-electron interactions. As discussed in Refs. \cite{ren95} and \cite{kan97}, coupled fractional QH channels can result in such metallic behaviour, as long as electrons (not fractional quasiparticles) backscatter the charge between the counterpropagating $\nu$ = 1/3 edges, creating an `antiwire'.  The $T$-dependence is predicted to be metallic since low-temperature correlations at $\nu$ = 1/3 suppress electron tunnelling and therefore backscattering.  The conductance is predicted to behave as a power-law $G(T) \sim T^\alpha$, and the data of Fig.~\ref{fig:Pdep} would fit an exponent $\alpha$ = -0.4, corresponding to the Luttinger parameter $g = 1 - \alpha/2 = 1.2$ after Ref. \cite{kan97}. We note that the same exponent occurs in the voltage dependence $dI/dV \sim V^\alpha$ [dashed line, Figs.~\ref{fig:Pdep}(a) and (b)].  If this explanation is relevant, it would appear that the accumulation wire effectively functions as a `vacuum' for counterpropagating fractional quasiparticles, permitting only electrons to backscatter.  We remark that the planar antiwire geometry originally suggested in Refs. \cite{ren95}, \cite{kan97} and implemented in Ref. \cite{kan00} actually prohibits the desired strong coupling of fractional QH edges, since the intervening barrier exponentially suppresses tunnelling at high $B$ \cite{gra01}.  Only in the non-planar geometry introduced here can counterpropagating edge modes overlap sufficiently in real space that  strong-backscattering in the high-$B$ limit may occur.

We note that the sharp confinement potential will play a decisive role in modeling the junction conductance.  Experimentally, sharp edge potentials have been shown to eliminate the incompressible strips characteristic of soft QH edges \cite{hub05}.  Recent theory has been able to describe conduction in this sharp limit where these incompressible strips are expected to be absent \cite{afif}.

In conclusion, we have characterized a new low-dimensional system, the bent QH junction.  Hartree calculations illustrate the dispersion in the junction, and show how non-planar confinement differs from planar.  The temperature and voltage dependence of the junction conductivity change with $\nu$, revealing metallic, weakly insulating, and strongly insulating states.  The length dependence reveals the influence of disorder.  The observation metallic behavior at a sharp junction may highlight the importance of interactions.

\begin{acknowledgments}
This work was supported by DFG Quanten-Hall-Schwerpunkt-Programm and EEC funding Contract No. RITA-CT-2003-505474.  M.G. is grateful to the von Humboldt Foundation for additional support, and also thanks T. Giamarchi, W. Kang, Eun-Ah Kim, A. MacDonald, D. Polyakov, I. Safi, and A. Siddiki for illuminating conversations, Joel Moser for the SEM picture, and Martin Geisler for measurement expertise.
\end{acknowledgments}


\begin{thebibliography}{10}

\bibitem{yac97}
A. Yacoby, H.~L. Stormer, K.~W. Baldwin, L.~N. Pfeiffer and K.~W. West, 
Solid State Comm. \textbf{101}, 77 (1997).

\bibitem{kau99}
D. Kaufman, Y. Berk, B. Dwir, A. Rudra, A. Palevski and E. Kapon, 
Phys. Rev. \textbf{B 59}, R10433 (1999).

\bibitem{fie90}
S.~B. Field, M.~A. Kastner, U. Meirav, J.~H.~F. Scott-Thomas, D.~A. Antoniadis, H.~I. Smith and S.~J. Wind, 
Phys. Rev. \textbf{B 42}, 3523 (1990).

\bibitem{tar95}
S. Tarucha, T. Honda and T. Sako,
Solid State Comm. \textbf{94}, 413 (1995).

\bibitem{hal82}
B. I. Halperin, 
Phys. Rev. \textbf{B 25}, 2185 (1982).  

\bibitem{wen90}
X.-G. Wen, 
Phys. Rev. Lett. \textbf{64}, 2206 (1990).

\bibitem{rod03}
S. Roddaro, V. Pellegrini, F. Beltram, G. Biasiol, L. Sorba, R. Raimondi and G. Vignale,
Phys. Rev. Lett. \textbf{90}, 046805 (2003); 
S. Roddaro, V. Pellegrini, F. Beltram, G. Biasiol and L. Sorba, 
Phys. Rev. Lett. \textbf{93}, 046801 (2004).

\bibitem{ji03}
Y. Ji, Y. Chung, D. Sprinzak, M. Heiblum, D. Mahalu and H. Shtrikman, 
Nature \textbf{422}, 415 (2003).

\bibitem{hau88}
R.~J. Haug, A.~H. MacDonald, P. Streda and K.~von Klitzing, 
Phys. Rev. Lett. \textbf{61}, 2797 (1988). 

\bibitem{chk92}
D.~B. Chklovskii, B.~I. Shklovskii and L.~I. Glazman, 
Phys. Rev. \textbf{B 46}, 4026 (1992).

\bibitem{cha94}
C. de C. Chamon and X.-G. Wen,
Phys. Rev. \textbf{B 49}, 8227 (1994).

\bibitem{kan00}
W. Kang, H.~L. Stormer, L.~N. Pfeiffer, K.~W. Baldwin and K.~W. West,  
Nature \textbf{403}, 59 (2000).

\bibitem{gra05}
M. Grayson, D. Schuh, M. Huber, M. Bichler and G. Abstreiter,
Appl. Phys. Lett. \textbf{86}, 032101 (2005).

\bibitem{sample}
Growth protocol for Sample A is 05-23-02.4L3a, and sample B is 01-30-03.2L3a.

\bibitem{hub05}
M. Huber, M. Grayson, M. Rother, W. Biberacher, W. Wegscheider and G. Abstreiter
Phys. Rev. Lett. \textbf{94}, 016805 (2005).

\bibitem{ren95}
S. R. Renn and D. P. Arovas,
Phys. Rev. \textbf{B 51}, 16832 (1995).

\bibitem{kan97}
C.~L. Kane and M.~P.~A. Fisher,
Phys. Rev. \textbf{B 56}, 15231 (1997).

\bibitem{but88}
M. Buttiker,
Phys. Rev. \textbf{B 38}, 9375 (1988).

\bibitem{abr79}
E. Abrahams, P.~W. Anderson, D.~C. Licciardello and T.~V. Ramakrishnan, 
Phys. Rev. Lett. \textbf{42}, 673 (1979).

\bibitem{gia88}
T. Giamarchi and H.~J. Schulz, 
Phys. Rev. \textbf{B 37}, 325 (1988).

\bibitem{gor05}
I.~V. Gornyi, A.~D. Mirlin and D.~G. Polyakov, 
Phys. Rev. Lett. \textbf{95}, 206603 (2005).


\bibitem{gra01}
M. Grayson, D.~C. Tsui, L.~N. Pfeiffer, K.~W. West and A.~M. Chang,
Phys. Rev. Lett. \textbf{86}, 2645 (2001).

\bibitem{wee88}
B.~J. van Wees, L. P. Kouwenhoven, H. van Houten, C.~W.~J. Beenakker, J.~E. Mooij, C.~T. Foxon and J. Harris,
Phys. Rev. \textbf{B 38}, 3625 (1988). 

\bibitem{afif}
A. Siddiki, and R. R. Gerhardts,
Phys. Rev. \textbf{B 70}, 195335 (2004).

\end{thebibliography}
\end{document}